\documentclass[oldversion]{aa}  
\usepackage{natbib}   
\bibpunct{(}{)}{;}{a}{}{,} 

\usepackage{graphicx}
\usepackage{txfonts}
\begin{document}
   \title{Properties of the molecular gas in a starbursting QSO at z$=1.83$ in the COSMOS field}


   \author{M. Aravena
          \inst{1,2}
          \and
          F. Bertoldi\inst{1} \and E. Schinnerer\inst{3} \and A. Weiss \inst{2} \and K. Jahnke\inst{3} \and C. L. Carilli\inst{4} \and D. Frayer\inst{5} \and C. Henkel \inst{2} \and M. Brusa\inst{6} \and K. ~M. Menten \inst{2} \and M. Salvato\inst{7} \and V. Smolcic\inst{7}}


   \offprints{M. Aravena}
\institute{
  Argelander Institut f\"ur Astronomie, Universit\"at Bonn, Auf dem H\"ugel 71, 53121 Bonn, Germany,\\
  \email{maraven@astro.uni-bonn.de}
  \and
  Max-Planck-Institut f\"ur Radioastronomie, Auf dem H\"ugel 69, 53121 Bonn, Germany
  \and
  Max-Planck-Institut f\"ur Astronomie, K\"onigsstuhl 17, 69117 Heidelberg, Germany
  \and
  National Radio Astronomy Observatory, P.O. Box, Socorro, NM 87801, USA 
\and
  Infrared Processing and Analysis Center, California Institute of Technology 100-22, Pasadena, CA 91125, USA
\and
  Max-Planck Institut f\"ur extraterrestrische Physik, Giessenbachstra{\ss}e 1, D-85748 Garching, Germany 
  \and
  California Institute of Technology 105-24, Pasadena, CA 91125, USA
%
%
  }
\authorrunning{Aravena et al.}
\titlerunning{Properties of Molecular Gas in a Composite Starburst-QSO}
   \date{Received ; accepted }

 
  \abstract
{Using the IRAM 30m telescope, we have detected the $^{12}$CO $J=2-1$, $4-3$, $5-4$, and $6-5$ emission lines in the millimeter-bright, blank-field selected AGN COSMOS J100038+020822 at redshift $z=1.8275$. The sub-local thermodynamic equilibrium (LTE) excitation of the $J=4$ level implies that the gas is less excited than that in typical nearby starburst galaxies such as NGC253, and in the high-redshift quasars studied to date, such as J1148+5251 or BR1202-0725. Large velocity gradient (LVG) modeling of the CO line spectral energy distribution (CO SED; flux density vs. rotational quantum number) yields H$_{2}$ densities in the range $10^{3.5}-10^{4.0}$ cm$^{-3}$, and kinetic temperatures between 50 K and 200 K. The H$_{2}$ mass of $(3.6 - 5.4) \times 10^{10}$ M$_{\sun}$ implied by the line intensities compares well with our estimate of the dynamical mass within the inner 1.5 kpc of the object. 
Fitting a two-component gray body spectrum, we find a dust mass of $1.2 \times 10^{9}$ M$_{\sun}$, and cold and hot dust temperatures of 42$\pm$5 K and 160$\pm$25 K, respectively.
The broad MgII line allows us to estimate the mass of the central black hole as $1.7 \times 10^{9}$ M$_{\sun}$. 
Although the optical spectrum and multi-wavelength SED matches those of an average QSO, the molecular gas content and dust properties resemble those of known submillimeter galaxies (SMGs).   The optical morphology of this source shows tidal tails that suggest a recent interaction or merger. Since it shares properties of both starburst and AGN, this object appears to be in a transition from a strongly starforming submillimeter galaxy to a QSO. 
}{}

\keywords{Galaxies: evolution -- Galaxies: starburst -- Galaxies individual (J100038+020822) -- Galaxies: ISM  -- Galaxies: quasars: emission lines -- Galaxies: high-redshift}
\maketitle
%

\section{Introduction}
\label{introduction}
Submillimeter blank field surveys have discovered a population of dust enshrouded high-redshift galaxies \citep{Smail1997,Hughes1998,Eales1999}, which are massive systems with huge molecular gas reservoirs and star formation at high rates \citep{Neri2003,Greve2005,Solomon2005}. Their faint X-ray emission \citep{Alexander2005} suggests that most of the submillimeter output is not powered by active galactic nuclei (AGN), but by massive star formation. These starbursting submillimeter galaxies (SMGs) account for a substancial fraction of the far-infrared (IR) background, and current models suggest this population may represent the formation of massive spheroidals at high-redshift \citep{Dunlop2001}. 

In the local Universe, an evolutionary connection between starbursting ultraluminous infrared galaxies (ULIRGs) and QSOs has been suggested \citep{Sanders1988a} and discussed controversially \citep[e.g.][]{Sanders1988a,Sanders1988b, Genzel1998, Tacconi2002}. This evolutionary cycle has found support from hydrodynamical simulations of galaxy formation \citep[e.g.][]{DiMatteo2005, Hopkins2005, Hopkins2006} and observations \citep{Sanders1988a,Page2004,Stevens2005}. 

The ubiquitous presence of AGN activity in most SMGs \citep{Alexander2005} suggests a close link between the AGN and starburst activity at redshifts $z \sim 1-3$. If a QSO is found to be far-IR luminous, it means that large amounts of gas and dust should still be present. Such a source may, in fact, be a good candidate for an object in the transition from a starburst to a QSO, in particular when it shows absorbed X-ray emission \citep{Page2004, Stevens2005}. Yet only a few high-redshift composite starburst/AGN have been studied in its molecular and multi-wavelength continuum emission \citep{RowanRobinson2000, leFloch2007, Coppin2008}. 

Observations of carbon monoxide (CO) in galaxies are important probes of the physical conditions of the cold and warm molecular gas in the galactic nuclei and disks. They provide estimates of the total amount of gas available to fuel starburst and/or AGN activity, and the CO line profile and intensity can be used to obtain important information about the galaxy kinematics, such as dynamical mass or size of the emitting region \citep{Solomon1997,Solomon2005}.

Because of their diagnostic value, great efforts have been made to observe CO emission lines in SMGs. These studies have benefited from deep radio continuum imaging (e.g. VLA 1.4 GHz) to locate the SMG accurately \citep{Ivison2002,Ivison2005, Ivison2007}, and from the determination of optical spectroscopic redshifts \citep[e.g.][]{Chapman2005}. Due the small spectroscopic bandwidths of current (sub)millimeter telescopes and interferometers, this was necessary to permit the proper frequency tuning for line observations (e.g. of CO, [CI], [CII]). Yet only 19 SMGs have been reported in CO so far \citep[][]{Frayer1998, Frayer1999, Andreani2000, Sheth2004, Hainline2004, Neri2003, Greve2005, Frayer2008, Coppin2008, Tacconi2008} and only a few have been observed in multiple molecular transitions to study the excitation conditions of their molecular gas reservoir \citep{Solomon2005}. Spatial structure and dynamics were studied for some SMGs through high resolution CO imaging \citep{Tacconi2006, Tacconi2008}. Overall, only $\sim 50$ high-redshift ($z>1$) objects have been detected in CO, most of which are luminous, optically selected QSOs \citep[][]{Omont1996, Guilloteau1997, Guilloteau1999, Carilli2002, Walter2003, Bertoldi2003, Beelen2004, Riechers2006, Carilli2007} and high-redshift radio galaxies \citep[HzRG;][]{DeBreuck2003, DeBreuck2005, Klamer2005, Papadopoulos2005}.

\begin{figure*}[!ht]
   \centering
   \includegraphics[width=15cm]{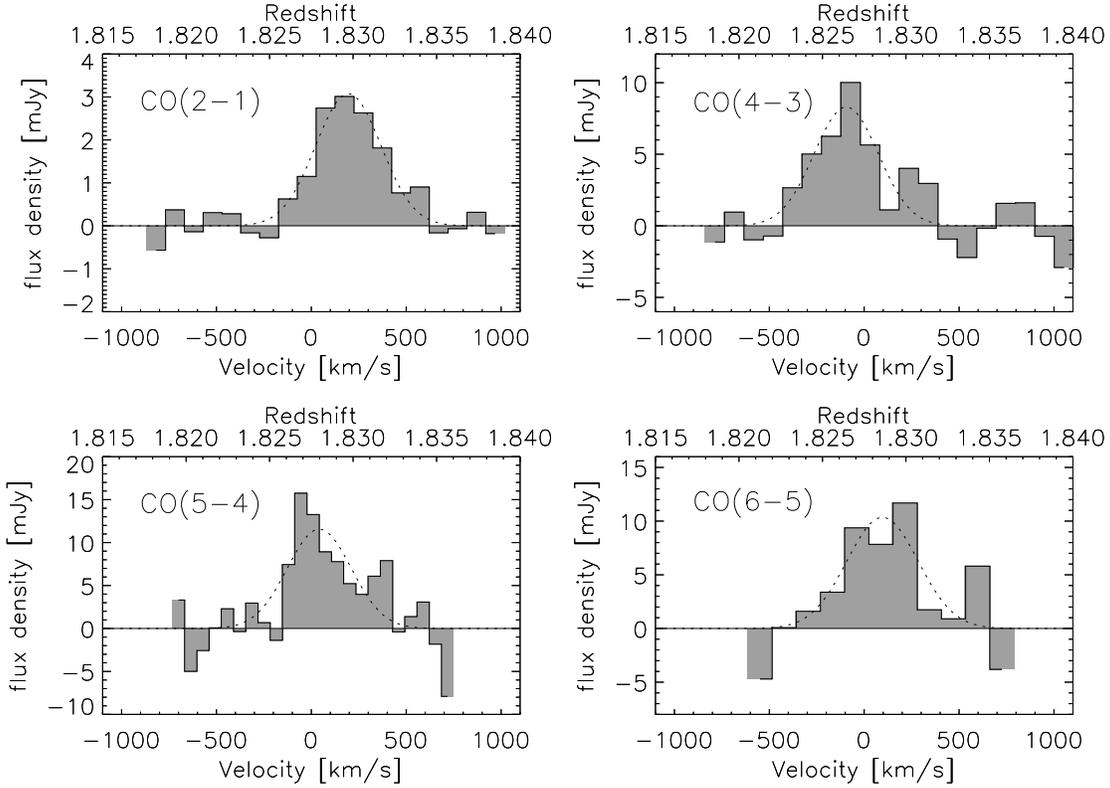}
      \caption{Observed spectra of the CO $J=2-1, 4-3, 5-4$ and $6-5$ emission lines in J100038+020822, obtained with the IRAM 30m telescope. Gaussian fits to the spectra are shown as dotted lines. The velocity reference is at $z=1.8275$.}
         \label{fig:lines}
\end{figure*}

Here we report the detection of CO $2-1$, $4-3$, $5-4$, and $6-5$ line emission from a millimeter selected QSO, J$100038.01+020822.4$. Its CO line intensities and flux density ratios allow us to estimate its molecular gas content and its excitation conditions. Its spectral properties suggest that this object may be evolving from a starburst to a QSO. 

After a description of the observations in Section 2, we present the molecular gas and dust properties of our source in Sections \ref{sect:lvg} to \ref{sect:co_prop2}. We study its morphology and multi-wavelength properties in Sections \ref{sect:optical_morph}, \ref{sect:sed} and \ref{sect:bhmass}. We discuss the results in Section 4 and give a brief summary in Section 5. Throughout, we use a $\Lambda$CDM cosmology, $H_{0}=70$ km s$^{-1}$ Mpc$^{-1}$, $\Omega_{\Lambda}=0.7$ and $\Omega_{\mathrm{M}}=0.3$.

\section{Observations}
\subsection{Source selection}
Pan-chromatic surveys are essential for an understanding of galaxy properties and their evolution. The cosmic evolution survey (COSMOS) is the first multi-wavelength survey to cover a sufficiently large area ($1.4\degr \times 1.4\degr$) at appropriate depth over nearly the entire electromagnetic spectrum to provide a comprehensive view of galaxy formation and large scale structure \citep[see][]{Scoville2007a}. 

As part of the COSMOS project, \citet{Bertoldi2007} mapped the central $\sim 20\arcmin \times 20\arcmin$ of the COSMOS field at 1.2 mm (250 GHz) using the Max-Planck millimeter bolometer camera (MAMBO) at the IRAM 30m telescope. An accompanying VLA radio imaging project \citep[1.4 GHz to $\sim 10\ \mu$Jy rms;][]{Schinnerer2007} allowed for the identification of radio counterparts for 24 millimeter sources.

The MAMBO source J100038.01+020822.4 (hereafter: J100038) is one of the strongest millimeter selected sources in COSMOS (boosting corrected $S_{1.2\ \mathrm{mm}} = 4.6 \pm 0.9$ mJy).

It hosts an X-ray luminous ($L_{\mathrm{X}} \sim 10^{44}$ erg s$^{-1}$), absorbed (log$N(\mathrm{H}) \sim 22-23$ cm$^{-2}$) AGN that classifies it as an obscured QSO \citep{Brusa2007,Mainieri2007}, and its optical spectrum is typical of a broad line (BL) AGN \citep{Trump2007}. The relatively faint 1.4 GHz radio emission ($S_{1.4 \mathrm{GHz}} = 237 \pm 27\ \mu$Jy) suggests that the millimeter emission arises from a starburst. In fact, the radio-to-millimetre flux ratio can be used as a redshift indicator \citep{CarilliYun1999} that implies $z=1.9$, which is consistent with the spectroscopic redshift. Since this is a millimeter selected QSO that has been detected in X-rays \citep{Stevens2005}, it likely constitutes a transitional case from SMG to QSO.

Three independent spectroscopic redshift measurements exist for J100038: $1.8325 \pm 0.0023$ \citep{Trump2007}, $1.825 \pm 0.002$ \citep{Prescott2006} and $1.8289 \pm 0.002$ (Marco Scodeggio, priv. comm.), which we have averaged to $z=1.8288 \pm 0.0037$.

\subsection{CO observations}

The CO observations were performed with the IRAM 30m telescope on Pico Veleta, Spain, during the winter 2006/2007 in good millimeter weather conditions (precipitable water vapor $\la 5$ mm). 
We observed the CO $2-1$ (redshifted to 81.551 GHz) and CO $5-4$ (203.850 GHz) lines simultaneously using the A$/$B receiver configuration, and the CO $4-3$ (163.088 GHz) and CO $6-5$ (244.603 GHz) lines using the C$/$D receiver configuration.

The observations were centered on the optical Subaru $I$ band position of J100038 at 10$^{h}$ 00$^{m}$ 38.01$^{s}$\ $+02^{\circ}$\ $08'$\ $22.6''$ (J2000). The optical position lies 0.04\arcsec \ away from the VLA 1.4 GHz position. The beam size of the IRAM 30m telescope at 200 GHz and 81 GHz is 12\arcsec \ and 30\arcsec, respectively.

We observed in wobbler switching mode with a wobbler rate of $0.5$ Hz and a wobbler throw of 60\arcsec \ in azimuth. Due to its proximity to the source, we used Saturn as main pointing and focus calibrator. The pointing was checked every hour and was stable within 3\arcsec \ during all runs. 

Typical values for the system temperatures were $\sim$ 140\ K, 400\ K, 460\ K and 600\ K for the 3 mm, 2 mm, and lower and higher 1 mm bands, respectively.  
We calibrated every 12 minutes with hot/cold absorbers and estimate the fluxes to be accurate to $\pm$10\% at 3 and 2 mm and 20\% at 1 mm. As spectrometers we used the 512 $\times$ 1 MHz filterbanks for the 3 mm receivers, and the 256 $\times$ 4 MHz filterbanks for the 2 and 1 mm receivers (1 GHz bandwidth). We reduced the data using CLASS, removing scans with strongly distorted baselines, substracted linear baselines in the remaining scans and rebinned the averaged spectra to velocity resolutions of 100, 120, 90 and 130 km\ s$^{-1}$ for the CO $2-1$, $4-3$, $5-4$ and $6-5$ lines, respectively. This led to baseline antenna temperature ($T_{\mathrm{A}}^{\ast}$) rms noise levels of 0.04, 0.17, 0.44 and 0.4 mK. To convert to main beam temperatures, $T_{\mathrm{mb}}$, we multiply $T_{\mathrm{A}}^{\ast}$ by the ratio between the forward and beam efficiencies, $F_\mathrm{eff}/B_\mathrm{eff}$, at the observed frequencies. Flux densities were obtained using the conversion factor $S_{\nu}/T_\mathrm{mb}=4.95$ Jy/K for the IRAM 30m telescope\footnote{\texttt{http://www.iram.es/IRAMES/telescope.html}}.

\begin{table*}
\caption{Observed line parameters for J100038.}             
\label{table:1}      
\centering                          
\begin{tabular}{l c c c c c c c c c c}        
\hline\hline                 
Transition~~~~~~~~~~~~~~~~~~ & $\nu_\mathrm{obs}$ $^{a}$ & HPBW $^{b}$& $T_{\mathrm{A}}$* $^{c}$& $T_{\mathrm{mb}}$ $^{d}$& $S_{\nu}$ $^{e}$& v$_{\mathrm{FWHM}}$ $^{f}$ & $I_{\mathrm{CO}}$ $^{g}$& $L'_{\mathrm{CO}}$ $^{h}$& $L_{\mathrm{CO}}$ $^{i}$\\    
           &  [GHz]  & [$\arcsec$]  & [mK] & [mK]    &   [mJy]   &  [km\ s$^{-1}$]  & [Jy\ km\ s$^{-1}$] & 10$^{10}$\ [K\ km\ s$^{-1}$\ pc$^{2}$] &  10$^{8}$\ [L$_{\sun}$]\\
\hline                        
   CO\ $2-1$ & 81.551 & 30.2 & 0.51 & 0.62  & 3.08 & $406 \pm 31$  & 1.33 & $5.6 \pm 0.35$ & $0.22 \pm 0.01$\\      
   CO\ $4-3$ & 163.088 & 15.1 & 1.18   & 1.69 & 8.39 &$427 \pm 73$ & 3.81 & $4.0 \pm 0.51$& $1.25 \pm 0.16$\\
   CO\ $5-4$ & 203.850 & 12.1 & 1.46   & 2.33 & 11.56 &$397 \pm 161$ & 4.87 & $3.3 \pm 0.8$& $2.00 \pm 0.49$\\
   CO\ $6-5$ & 244.603 & 10 & 1.19& 2.09  & 10.35 &$443 \pm 75$ & 4.86 & $2.3 \pm 0.4$& $2.40 \pm 0.47$\\
   \hline                                   
\end{tabular}
\begin{flushleft}
\noindent {\it Notes:} $^{a}$ Observed frequency; $^{b}$ Half power beam width of the IRAM 30m telescope at the observed frequency; $^{c}$ Antenna temperature; $^{d}$ Main beam temperature; $^{e}$ Flux density; $^{f}$ CO line full width half maximum; $^{g}$ Integrated CO intensity ; $^{h}$ CO luminosity; $^{i}$ Integrated CO luminosity.
\end{flushleft}
\end{table*}

\section{Results and Analysis}

Figure \ref{fig:lines} shows the resulting spectra and Table \ref{table:1} summarizes the measured line parameters.
The line profiles for the different transitions appear Gaussian and similar to each other, with no signs of velocity structure,  as might be expected for a merger and is often found for SMGs \citep{Greve2005, Weiss2005b}. However, the signal to noise ratio is too low to examine the individual line profiles. Although the CO $4-3$ line appears to be slightly shifted in velocity relative to the other lines, the line widths are very similar, with an average of $\Delta \mathrm{v}_{\mathrm{FWHM}} = 417 \pm 48$ km s$^{-1}$. The spectral energy distribution of the CO rotational emission (CO SED) peaks at the $J=5-4$ transition (Fig. 2 and Table \ref{table:1}). Using the detected CO lines we estimate a systemic CO redshift $z=1.8275\pm0.0013$, which we use throughout this paper.

\subsection{Large Velocity Gradient model}
\label{sect:lvg}
\begin{figure}[t]
   \centering
   \includegraphics[scale=0.55]{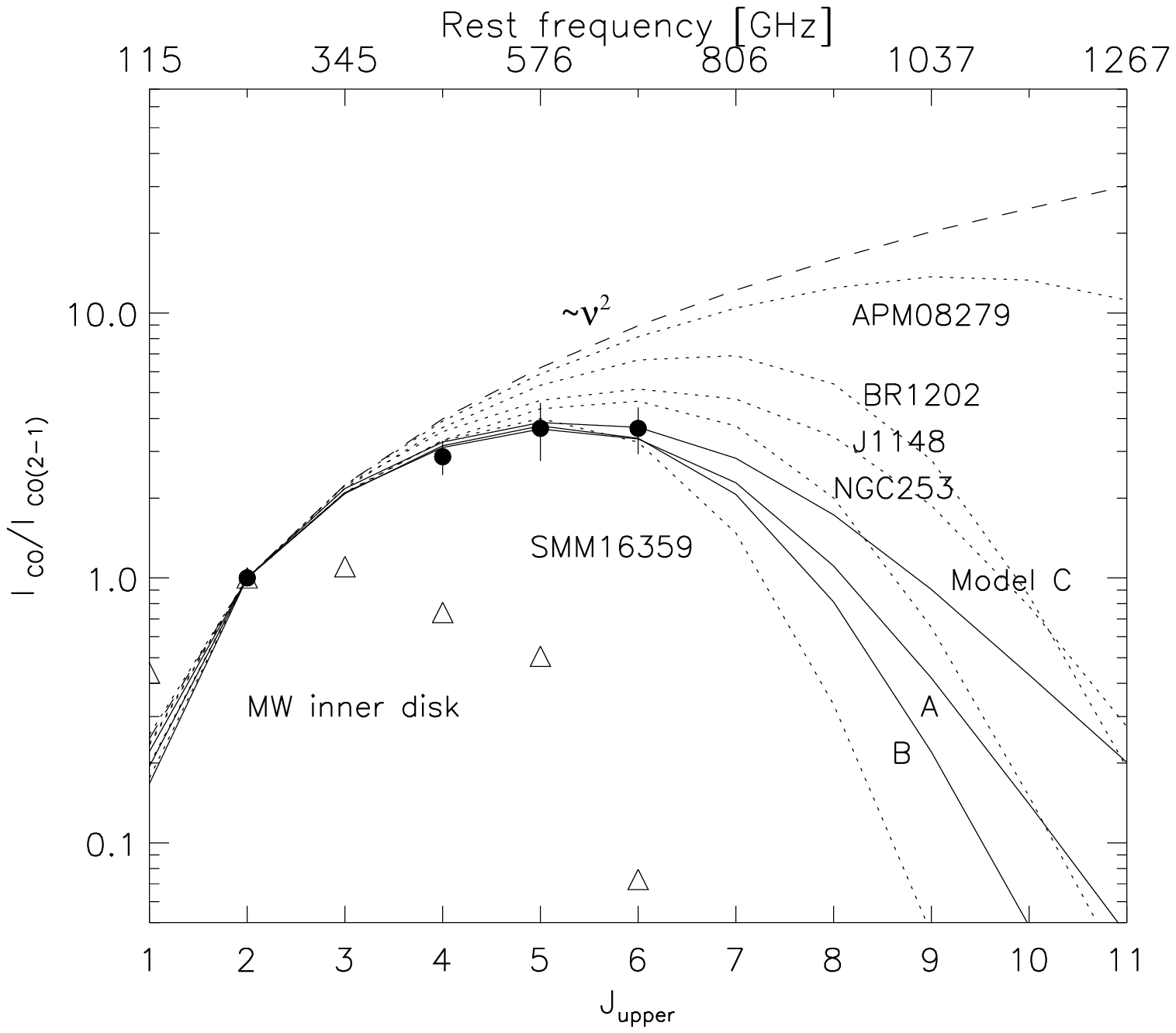}
      \caption{CO line SED: integrated line flux, $I_{\mathrm{CO}}$, normalized to CO $2-1$, vs. rotational quantum number. Filled squares mark the integrated line fluxes for J100038. 
Solid lines show LVG models A, B and C (see text). Dotted lines show the CO line SEDs for NGC253 \citep{Guesten2006}, SMM16359 \citep{Weiss2005b} and APM08279 \citep{Weiss2007}. Open triangles show the CO SED for the inner Milky Way \citep{Fixsen1999}. The dashed line shows the line flux increasing as $\nu^{2}$, which would be expected for optically thick LTE conditions.
              }
         \label{fig:lvg1}
\end{figure}

\begin{figure}[t]
   \centering
   \includegraphics[scale=0.55]{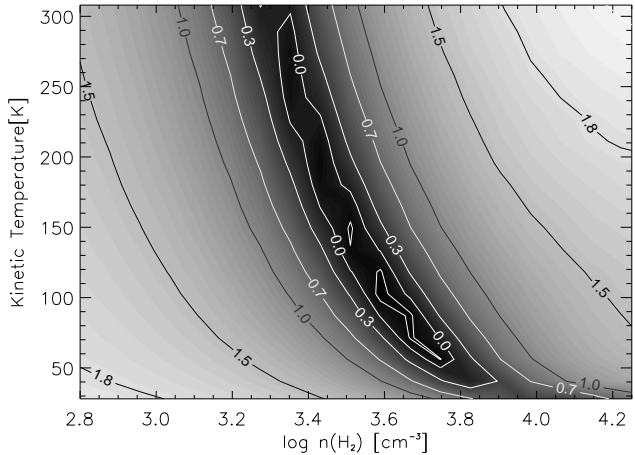}
      \caption{log$\chi^{2}$ distribution for the single component LVG model as a function of $T_{\mathrm{kin}}$ and $n(\mathrm{H}_{2})$ with $[\mathrm{CO}]/\mathrm{(dv/dr)} = 10^{-5}$ pc (km\ s$^{-1}$)$^{-1}$. Best fits are provided for solutions with $\mathrm{log}_{10} \chi^{2}< 0$.
              }
         \label{fig:lvg2}
\end{figure}

To study the molecular gas excitation in J100038 we compared the relative CO line intensities with those predicted by a single component large velocity gradient (LVG) model, assuming spherical geometry. We used the collision rates from \citet{Flower2001} with an ortho-para H$_{2}$ ratio of 3 and a fixed CO abundance per velocity gradient of $[\mathrm{CO}]/\mathrm{(dv/dr)} = 10^{-5}$ pc (km\ s$^{-1}$)$^{-1}$ \citep[e.g.][]{Weiss2005a,Weiss2007}.

The best fit to the data is provided by a model with a kinetic temperature, $T_{\mathrm{kin}}$ of $ 95$ K and an H$_{2}$ density, $n(\mathrm{H}_{2})$ of $10^{3.6}$ cm$^{-3}$ (model A, Fig. \ref{fig:lvg1}). Similarly good fits are achieved with $T_{\mathrm{kin}}$ $= 60$ K and $n(\mathrm{H}_{2}$) $= 10^{3.7}$ cm$^{-3}$ (model B), or $T_{\mathrm{kin}}$ $= 200$ K and $n(\mathrm{H}_{2}$) $= 10^{3.5}$ cm$^{-3}$ (model C). They all show a peak in the CO SED at the $J=5-4$ line, but predict different higher-$J$ intensities. 

Figure \ref{fig:lvg2} shows the $\mathrm{log}_{10} \chi^{2}$ distribution as a function of the kinetic temperature and H$_{2}$ density. The temperature--density parameter space is degenerate at densities between $10^{3.0}$ cm$^{-3}$ and $10^{4.0}$ cm$^{-3}$. The most likely ranges for the kinetic temperature and the H$_{2}$ density are $\sim 50$ to $200$ K and $10^{3.5}$ to $10^{4.0}$ cm$^{-3}$, respectively.

The CO abundance per velocity gradient relates to the line opacities predicted by the LVG models \citep{Weiss2007}. Higher values of $[\mathrm{CO}]/\mathrm{(dv/dr)}$ correspond to higher opacities for the high-$J$ transitions. From Figure \ref{fig:lvg1} we see that all models predict optically thick conditions and thermalized emission in the $J_{\mathrm{upper}}<4$ lines, consistent with the line SED following $\nu^2$ and supporting our CO column per velocity gradient selection.

Although the kinetic temperature is poorly constrained, the LVG modelling prediction for the CO $1-0$ line intensity does not strongly depend on the model chosen:
$I_{\mathrm{CO}\ 1-0} = 0.27 - 0.4$ Jy km s$^{-1}$.
 
\begin{figure}[t]
   \centering
   \includegraphics[scale=0.55]{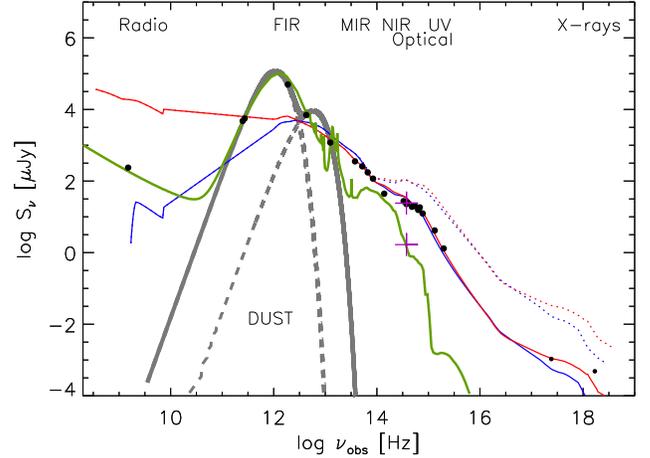}
      \caption{Spectral energy distribution (SED) for J100038. The photometric data points (filled circles) range from radio waves to X-rays. The lower purple cross shows the ACS $I$ band flux derived for the host galaxy while the upper one shows the flux for the nucleus. The red and blue curves are the average radio-loud and radio-quiet QSO SEDs from \citet{Elvis1994}. The dotted lines show the \citet{Elvis1994} models, not corrected for extinction, and the solid lines show the SEDs corrected with the \citet{Calzetti2000} extinction model. The green line shows the SED of the typical starburst galaxy Arp220. The thick gray lines show our two-component dust model. 
              }
         \label{fig:sed}
\end{figure}

\subsection{CO Size and Mass}
\label{sect:co_prop}
The ratio between the brightness temperatures predicted by the LVG models, $T_{\mathrm{b}}$, and the observed line temperatures, $T_{\mathrm{mb}}$, yields an estimate of the CO emitting region size. 
The angular source size $\theta_\mathrm{s}$ is related to the beam size $\theta_\mathrm{b}$ through
\begin{equation}
 \theta_{\mathrm{s}}=\theta_{\mathrm{b}}\times\left(\left[\frac{T_{\mathrm{b}}}{(1+z)T_{\mathrm{mb}}}\right]-1\right)^{-1/2}. 
\end{equation}

The source sizes we thereby compute for the different fit models (A, B and C) and line brightness temperatures agree well, $\theta_{\mathrm{s}} \approx 0.17-0.20 \arcsec$. 
At the source redshift ($z=1.8275$), 1\arcsec \ corresponds to $8.44\ \mathrm{kpc}$, so that the estimated source size lies in the range of $1.4 - 1.7$ kpc. 

The CO luminosity can be derived from the CO line intensity, following \citet{Solomon1997} as
\begin{equation}
L'_{\mathrm{CO}} = 3.25 \times 10^{7}\ S_{\mathrm{CO}}~\Delta{\rm v}~\nu_{\mathrm{obs}}^{2}\ D_{\mathrm{L}}^{2}\  (1+z)^{3} \rm ~~~K~ km~ s^{-1}~ pc^{2}
\end{equation}
where $S_{\mathrm{CO}} \Delta \rm v$ is the integrated CO line flux in Jy km s$^{-1}$, $D_{\mathrm{L}}$ is the luminosity distance in Mpc and $\nu_{\mathrm{obs}}$ is the observed line frequency in GHz. Our predicted CO flux $I_{1-0} = S_{1-0} \Delta V = 0.27 -0.4$ Jy km s$^{-1}$ yields $L'_{1-0} = (4.5 - 6.7) \times 10^{10}$ K\ km\ s$^{-1}$\ pc$^{2}$. 

The CO $1-0$ luminosity is commonly used to estimate the molecular gas mass. Adopting the conversion factor between CO luminosity and H$_{2}$ mass, $\alpha = 0.8\ M_{\sun}$ (K km s$^{-1}$ pc$^{2}$)$^{-1}$, that was derived for local ULIRGs \citep{Downes1998}, we find  an H$_{2}$ mass of $(3.6 - 5.4) \times 10^{10}$ M$_{\sun}$.

From the size estimate of the dense gas region and the average CO line FWHM, $\Delta {\rm v} = 417$ km s$^{-1}$, we can estimate a dynamical mass contained in the emitting region. For a radius of 0.75 kpc we infer a dynamical mass $M_{\mathrm{dyn}} = 3 \times 10^{10}$ sin$^{-2}$(i) $M_{\sun}$. For an average inclination angle $i = 30 \degr$,  this leads to $1.2 \times 10^{11} M_{\sun}$, which agrees well with the CO luminous mass estimate. Assuming a disk-like geometry would lower the dynamical mass estimate by a factor $2/\pi$.

The molecular gas mass can thus account for a large fraction ($\approx 30-40\%$) of the dynamical mass out to 0.75 kpc. This result suggests a somewhat larger fraction than in local ULIRGs in which the molecular gas mass represents 16\% of the dynamical mass in the nuclear regions \citep{Downes1998}.

SMGs and high-redshift QSOs have typical H$_{2}$ masses, $M(\mathrm{H}_{2})$ of $(3.0\pm1.2)\times10^{10}\ M_{\sun}$ within a 4 kpc diameter \citep{Greve2005, Solomon2005}, consistent with the value found in our source. Dynamical masses of SMGs and high-redshift QSOs range between 1.0 and 2.0$\times10^{11}\ M_{\sun}$ \citep{Greve2005,Tacconi2006,Solomon2005}, which also agrees well with the result exposed above. We note that our dynamical mass estimate has been computed assuming a radius of 0.75 kpc unlike the case of typical SMGs and high-redshift QSOs, for which it is computed with a radius of 2 kpc. Scaling to this radius, i.e. adopting a continuous CO distribution, we find that the dynamical mass for J100038 would be $1.6\times10^{11}\ M_{\sun}$, still in agreement with values for SMGs and high-redshift QSOs.

\subsection{Dust Continuum}
\label{sect:dust_prop}

\begin{table}
\caption{Infrared and radio fluxes of J100038. $^{\dagger}$}
\label{table:2}      
\centering                          
\begin{tabular}{l l c}        
\hline\hline 
Band & Flux density & Unit \\
\hline
IRAC 3.6 $\mu$m~~~~~~~~~~~~~~~~~~~~~~~~~~~~~~~~~~~~~~~ & $116.1 \pm 0.2$ &~~$\mu$Jy \\
IRAC 4.5 $\mu$m & $175 \pm 0.4$ &~~$\mu$Jy \\
IRAC 5.8 $\mu$m & $258.1 \pm 1.1$ &~~$\mu$Jy \\ 
IRAC 8.0 $\mu$m &  $356.4 \pm 2.3$ &~~$\mu$Jy\\
MIPS 24 $\mu$m & $1.43 \pm 0.1$ &~~mJy \\
MIPS 70 $\mu$m  & $7 \pm 2$  &~~mJy \\
MIPS 160 $\mu$m & $50 \pm 15$  &~~mJy \\
Bolocam 1.1 mm & $5.6 \pm 1.9$  &~~mJy \\
MAMBO 1.2 mm & $4.8 \pm 1.0$  &~~mJy\\
VLA 1.4 GHz & $237 \pm 37$  &~~$\mu$Jy\\   
\hline                                   
\end{tabular}
\begin{flushleft}
 \noindent $^{\dagger}$ IRAC and MIPS aperture corrected (total) flux densities are from the S-COSMOS data \citep[][]{Sanders2007}; Flux densities at 1.1 mm and 1.2 mm are from Aguirre et al. (2006) and
\citet{Bertoldi2007}, respectively; Radio flux is from VLA-COSMOS \citep{Schinnerer2007}.\\
\end{flushleft}

\end{table}

To study the dust properties of J100038 we use a $\chi^{2}$ minimization procedure to fit a 2-component gray-body spectrum to the 5 IR photometric data points from 24 $\mu$m to 1.2 mm (observed frame). A single component gray-body spectrum does not provide a good approximation to all points. We do not assume optically thin emission (dust optical depth, $\tau_{\nu} \ll 1$) but use the complete expression for the flux density \citep[see][]{Weiss2007}:
\begin{equation}
S_{\nu} = \frac{\Omega}{(1+z)^{3}}\left[ B_{\nu}(T_{\mathrm{dust}})-B_{\nu}(T_{\mathrm{BG}})\right] (1-e^{-\tau_{\nu}}),
\label{eq:flux}
\end{equation}
where $B_{\nu}(T)$ is the Planck function, $T_{\mathrm{dust}}$ is the dust temperature and $T_{\mathrm{BG}}$ is the cosmic background temperature at the source redshift, $T_{\mathrm{BG}}=2.73\times(1+z)$ K. We define the apparent solid angle subtended by the source as $\Omega=\pi(d_{0}/D_{\mathrm{A}})^{2}$, with $d_{0}$ being the equivalent source size which we assume to be 1.5 kpc based on the derived CO size, and $D_{\mathrm{A}}$ the angular distance at the source redshift. The dust optical depth is

\begin{equation}
\tau_{\nu}=\frac{\kappa(\nu)M_{\mathrm{dust}}}{D_{\mathrm{A}}^2\Omega},
\end{equation}
where $M_{\mathrm{dust}}$ refers to the dust mass, and the dust absorption coefficient has the form $\kappa(\nu) = \kappa_{0} (\nu/\nu_{0})^{\beta}$. We adopt an emissivity index $\beta = 2.0$ \citep{Priddey2001} and $\kappa_{0} = 0.4$ cm$^{2}$ g$^{-1}$ at 250 GHz \citep{Kruegel1994}.

The best fitting model leads to a dominant cold component with $T_{\mathrm{C}}=42\pm5$ K, and a hot component with $T_{\mathrm{H}}=160\pm25$ K. We calculate a dust mass, $M_{\mathrm{C}}$ for the cold component of $(1.2\pm0.4)\times10^{9}\ M_{\sun}$; however the mass implied for the hot component is only $M_{\mathrm{H}}=(4\pm2)\times10^{4}\ M_{\sun}$. Therefore, the cold dust component accounts for almost all the dust mass. We note that for a dust mass of $\sim1\cdot10^{9}\ M_{\sun}$ and a size of 1.5 kpc, the emission becomes optically thick ($\tau_{\nu}\geq1$) at $\lambda \leq 130\ \mu$m (rest frame), supporting our model selection.


The far-IR luminosity can be computed from the modeled flux density as

\begin{equation}
L_{\mathrm{FIR}} = \frac{4 \pi\ D_{\mathrm{A}}^{2}}{(1+z)^{3}} \int S_{\nu} d\nu.
\end{equation}

Integrating from 50 to 1000 $\mu$m \citep{Omont2001}, we find $L_{\mathrm{FIR}}=8.5\times10^{12}\ L_{\sun}$. Although the contribution from hot dust is only $L_{\mathrm{FIR,\ H}}=5.1\times10^{10}\ L_{\sun}$, most of it is provided by the cold dust component with $L_{\mathrm{FIR,\ C}}=8.4\times10^{12}\ L_{\sun}$. We note that the selection of different integration limits implies different values for the luminosity, in particular for the hot dust component. If we integrate from 8 to 1000 $\mu$m \citep{Sanders1996}, we obtain $L_\mathrm{FIR}=1.3\times10^{13}\ L_{\odot}$, $L_\mathrm{FIR,c}=9.4\times10^{13}\ L_{\odot}$ and $L_\mathrm{FIR,h}=3.7\times10^{12} L_{\odot}$ for the total, cold and hot component luminosities, respectively.

A substantial difference seems to exists in the dust temperatures observed in SMGs \citep[$T_{\mathrm{dust}}\sim$35 K;][]{Kovacs2006, Pope2006, Blain2004} and high-redshift QSOs \citep[$T_{\mathrm{dust}}\sim$45 K;][]{Omont2003, Beelen2006, Wang2008}. This difference could possibly be explained by the fact that starburst galaxies with warmer dust would have their dust emission shifted to shorter wavelengths making them more difficult to detect at submillimeter wavelengths \citep{Blain2004}. If this difference is real, it would imply that the dust properties of J100038 are more consistent with high-redshift QSOs.

\subsection{Star Formation Rates and Efficiencies}
\label{sect:co_prop2}

Assuming that the AGN does not contribute significantly to the heating of the cold and hot dust components and that most of the dust emission is produced by starburst activity, the far-IR luminosity implies an estimate for the star formation rate (SFR). \citet{Omont2001} found the following relation between the far-IR luminosity and the SFR,
\begin{equation}
 \mathrm{SFR}\ [M_{\sun}\ \mathrm{yr}^{-1}] = \delta \left( \frac{L_{\mathrm{FIR}}}{10^{10}\ L_{\sun}}\right) ,
\end{equation}
where $\delta$ is a function of the initial mass function (IMF) and vary in the range $0.8-3$. For $\delta=2$, we then find SFR $\approx1700\ M_{\sun}$ yr$^{-1}$, well in agreement with the value found using the IR based estimator from \citet{Kennicutt1998}, SFR $\approx2150 \ M_{\sun}$ yr$^{-1}$.


The ratio between the far-IR luminosity and the CO luminosity can be interpreted as a star formation efficiency, SFE $=L_{\mathrm{FIR}}/L'_{\mathrm{CO}} = 130 - 190\ $ $L_{\sun}$ (K km s$^{-1}$ pc$^{2}$)$^{-1}$, or in terms of the gas mass implied by the CO luminosity, SFE $= 160 - 235\ L_{\sun}\ M_{\sun}^{-1}$.

This value is similar to the average value found for local ULIRGs, $\left\langle \mathrm{SFE}\right\rangle = (180\pm160)\ L_{\sun}\ M_{\sun}^{-1}$ \citep{Solomon1997}, although it appears to be slightly lower than those obtained for SMGs, $\left\langle \mathrm{SFE}\right\rangle = 450\pm170\   L_{\sun}\ M_{\sun}^{-1}$ \citep{Greve2005}.

The luminosity ratio can also be interpreted as a gas depletion time,
$\tau_{\mathrm{SF}} = M(\mathrm{H}_{2})/\mathrm{SFR} \approx 5\times10^{10}\ M_{\odot}/1700\ M_{\odot}\ \mathrm{yr}^{-1} = 30$ Myr. Others \citep[e.g.][]{Greve2005} have estimated the gas depletion time in SMGs as $\sim 16$ Myr, with a range of $10 - 100$ Myr \citep{Solomon2005}, well in agreement with what we find for J100038, but shorter than that found in ULIRGs or normal spirals.

\subsection{Optical Morphology}
\label{sect:optical_morph}
\begin{figure}[!ht]
   \centering
   \includegraphics[scale=0.65]{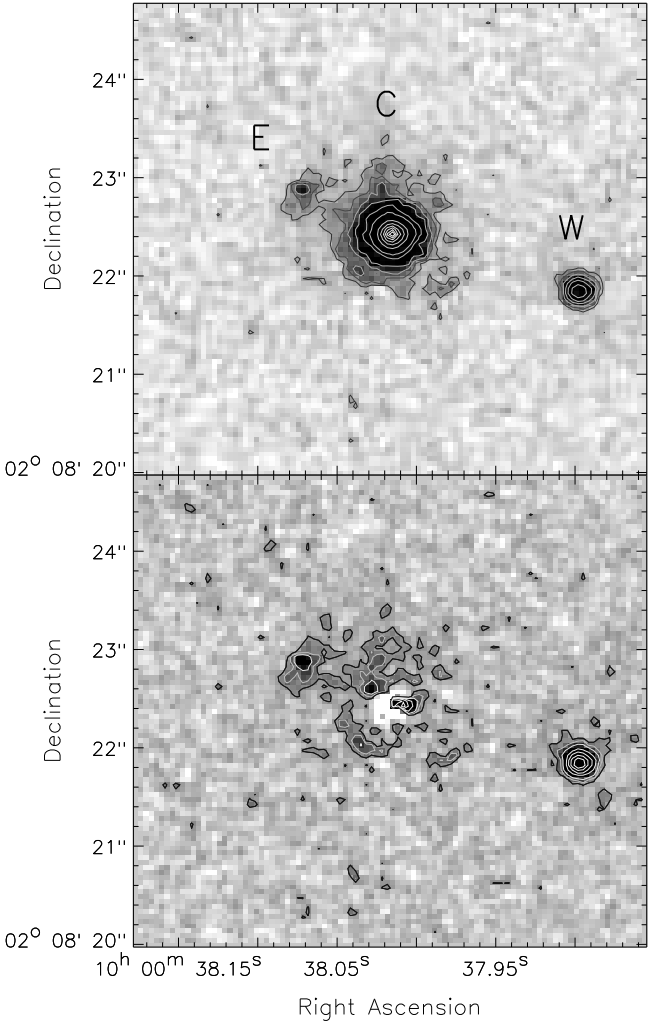}
      \caption{Optical morphology of J100038. The HST ACS $I$ band image ($\lesssim0.1\arcsec$ resolution) is shown with contours scaled in magnitude. The top panel shows the original image centered at the central component $C$, which coincides with the QSO and X-ray source. Two additional components are seen toward the east ($E$) and west ($W$). The bottom panel shows the nucleus substracted HST ACS $I$ image centered at the same position. In addition to the $E$ and $W$ components, patchy structure is seen in the central component. 
}
         \label{fig:closeup_i}
   \end{figure}

The Hubble Space Telescope Advanced Camera for Surveys (HST ACS) $I$ band image of J100038  (Figure \ref{fig:closeup_i}) shows a complex morphology with three discernable components. For the bright central component $C$,  Figures \ref{fig:sed} and \ref{fig:spectra} show the SED and optical spectrum, respectively.
 

The western component is offset by $\sim 2\arcsec$ from the central source and shows no sign of a connection to it.
Although the Subaru $I$ band images previously shown by \citet{Bertoldi2007} are only able to separate $C$ and $W$, the HST imaging interestingly shows a third eastern component ($E$).

Since the three components are well aligned, it is suggestive that $E$ and $W$ could be gravitationally lensed images. It could also be possible that all three components are lensed images as in e.g. APM08279 \citep{Weiss2007}.
However, the optical photometric redshift of $1.4-1.5$ for $W$, which is resolved in all optical images \citep[see][]{Bertoldi2007}, is consistent with that of C, with no significant secondary solution at higher redshift, thus ruling out the possibility that E and W is a background object lensed by C.
Imaging spectroscopy would much help to establish a dynamical relation between the three components, which could support a merger history as the cause for the starburst and nuclear activity.

\subsection{Spectral Energy Distribution}
\label{sect:sed}

Figure \ref{fig:sed} shows the J100038 spectral energy distribution from the radio regime to X-rays. For comparison, we show the (redshifted to $z=1.83$) SED of the proto-typical starburst galaxy Arp220 \citep{Silva1998}, which fits well from radio to IR. The match strongly supports the conclusion that most of the radio to far-IR emission of J100038 is produced by star formation. 

We also compare the observed photometry with the average QSO SEDs of \citet{Elvis1994}. While the near to mid-IR photometry of J100038 is well matched by both the radio loud and radio quiet QSO templates, at higher frequencies we need to apply a reddening correction to obtain a good fit. Although the radio-loud model would fit the millimeter flux, it is clearly inconsistent with the low observed radio flux. 
To match the model at the higher frequencies, we applied an extinction correction based on the model of \citet{Calzetti2000} for starburst galaxies.
A correction with $A_{\mathrm{V}} \approx 0.7$ yields a good match between the models and the UV/optical photometry, and for the radio-loud model even for the X-ray flux. Given the redshift ($z=1.8275$) and the best fit reddened template, we derive an extinction corrected absolute magnitude in the Subaru $V$ band, $M_{\mathrm{V}}$ of $-24.2$ magnitudes.



At near-IR and shorter wavelengths, the photometry corresponds to the central plus eastern component, but at longer wavelengths, the western component cannot be distinguished any longer.  
In the optical and radio regimes, the flux of component W is only 20\% of that of C, but differential extinction could in principle be the cause of this difference. 

Based on an empirical point-spread-function created from isolated stars in the COSMOS field in the vicinity of J100038, Jahnke et al. (in preparation) studied the contribution from the AGN and host galaxy to the optical emission in J100038 \citep[see also][]{Jahnke2004}. They find that the host galaxy is resolved and has irregular structure outside the central 0.5\arcsec, where nuclear residuals might dominate, as shown in Figure \ref{fig:closeup_i}.

The host galaxy contributes with a $\sim(8 \pm 2)$\% to the total source flux. Therefore, most of the emission at optical wavelengths ($92 \pm 1$\% of the total flux) arises from an unresolved point source, with $I_{\mathrm{ACS, host}}=23.07\pm0.1$ magnitudes for the host and $I_{\mathrm{ACS, nucleus}}=20.43\pm0.02$ magnitudes for the nucleus. This gives $I_{\mathrm{ACS,total}}=20.34\pm0.02$ magnitudes in total. Component E contributes a 0.8\% to the total nucleus$+$host flux, $I_{\mathrm{ACS, E}}=25.6\pm0.2$. 


Figure \ref{fig:sed} shows the ACS $I$ band fluxes for the host galaxy and nucleus. As mentioned above, the main flux contribution at this wavelength is provided by the latter. This is also true for the flux in the rest of the optical bands, however it is interesting to note that the $I$ band flux density of the host galaxy component matches the optical emission of the Arp220 SED, suggesting that it is heavily absorbed by dust and hosts starburst activity. 

\subsection{Black Hole Mass Estimate}
\label{sect:bhmass}

\begin{figure}[t]
   \centering
   \includegraphics[scale=0.5]{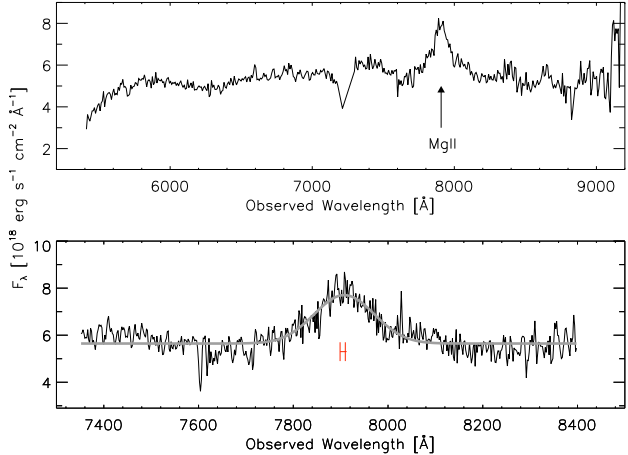}
      \caption{Top: optical Magellan spectrum of J100038 \citep{Trump2007}. The broad line at $\lambda \approx 7900$\ \AA \ is identified as MgII. Bottom: close-up of the MgII line. The smooth curve shows a Gaussian fit. The red horizontal error bar represents the position of the MgII line at the average CO redshift ($z=1.8275$) and CO line width ($\sim400$ km s$^{-1}$).
              }
         \label{fig:spectra}
\end{figure}

To estimate the mass of the central supermassive black hole one commonly assumes that the motion of the emitting gas around the black hole is virialized \citep{Peterson2000}. Then the black hole mass is related to the Keplerian velocity of the broad-line region (BLR) gas, $V_{\mathrm{BLR}}$, and to the BLR radius, $R_{\mathrm{BLR}}$, through $M_{\mathrm{BH}} \propto V_{\mathrm{BLR}}^{2} R_{\mathrm{BLR}}$. The BLR velocity is proportional to the line width of the MgII line (FWHM[MgII]) and the BLR radius has been observed to correlate with the UV continuum emission at 3000 \AA \ \citep{Mclure2002}. Combining these relations \citep[see ][]{Mclure2004},
\begin{equation}
\frac{M_{\mathrm{BH}}}{M_{\sun}} = 3.2\times \left( \frac{\lambda L_{3000}}{10^{37} \mathrm{W}}\right) ^{0.62} \left(\frac{\mathrm{FWHM}[\mathrm{MgII}]}{\mathrm{km}\ \mathrm{s}^{-1}}\right)^{2},
\end{equation}
where $\lambda L_{3000}$ is the luminosity at 3000 \AA \ (rest frame). In the optical spectrum of J100038 \citep{Trump2007}, the MgII line is prominent and allows for an estimate of the black hole mass (Figure \ref{fig:spectra}). Assuming a Gaussian shape for the MgII doublet we find a FWHM of $150.6 \pm 37.8$ \AA \ or $5714 \pm1442$ km\ s$^{-1}$. From the average radio-quiet QSO SED fit we estimate that the rest frame (extinction corrected) 3000 \AA \ luminosity is $8.71\times 10^{38}$ W, resulting in a black hole mass of $M_{\mathrm{BH}} = (1.7 \pm 0.8)\times 10^{9} M_{\sun}$. 

The maximum luminosity that can be reached by an accreting black hole of this mass, or Eddington limit, is $L_{\mathrm{Edd}}=5.6\times10^{13}\ L_{\sun}$. Assuming a typical QSO SED \citep{Elvis1994}, we obtain a bolometric luminosity of $L_{\mathrm{bol}}\approx 3\times10^{13}\ L_{\sun}=0.54\times L_{\mathrm{Edd}}$ which implies a moderate accretion rate (far from the Eddington limit).

\section{Discussion}
\label{sect:discussion}
\subsection{Comparison of Excitation Conditions}

The observations of the CO lines in J100038 show that its molecular gas is somewhat less excited than observed in local starburst or high-redshift QSOs. We find that the turn-over of the CO line SED occurs between the CO\ $5-4$ and CO\ $6-5$ transitions whereas in most local starbursts/AGNs, or high-redshift QSOs studied to date, the peak of the CO line SED is typically located between the CO\ $6-5$ and CO\ $7-6$ transitions. Examples of such higher excitation CO emission in the local Universe are NGC253 \citep{Guesten2006}, M82 \citep{Weiss2005a}, and at high-redshift, J1148+5251 \citep{Bertoldi2003,Walter2003}, BR1202-0725 \citep{Carilli2002, Riechers2006} and APM08279+5255 \citep{Weiss2007}. The only SMG for which the CO SED has been traced over its peak is J16359+6612 \citep{Weiss2005b}, which interestingly, shows a similar CO excitation to J100038. Other examples of lower excitation are found toward the centers of the local starburst/AGN of Circinus and NGC4945 \citep{Hitschfeld2008}, or in the main starburst region of the Antennae galaxy \citep{Zhu2003}. The lower excitation of the molecular gas observed in these cases, and in particular towards J100038, is likely produced by the relatively low H$_{2}$ density values ($n(\mathrm{H}_{2})<10^{4}$ cm$^{3}$). Furthermore, the best solution derived from the LVG analysis indicates a moderate kinetic temperature ($\sim95$ K), suggesting that the AGN does not contribute strongly to the heating of the molecular gas. Conversely, this may imply that most of the heating is produced by star formation, as it is also suggested by the similarity of molecular gas conditions ($T_{\mathrm{kin}}$, $n$(H$_{2}$)) observed between J16359+6612 and J100038.

\begin{table}
\caption{Summary of derived properties of J100038+020822}             
\label{table:3}      
\centering
\begin{tabular}{l l c}        
\hline\hline
Property & Name & Value \\
\hline 
$M(\mathrm{H}_{2})$ &Molecular Gas Mass ($10^{10}\ M_{\sun})$ &  $3.6 - 5.4$  \\
$M_{\mathrm{dyn}}$  &Dynamical Mass ($10^{10}\ \mathrm{sin}^{-2}(i)\ M_{\sun}$)& $3.0$  \\
$M_{\mathrm{dust}}$             &Dust Mass ($10^{9}\ M_{\sun}$)  &  $1.2$\\
SFR                 &Star Formation Rate ($10^{3}\ M_{\sun}$ yr$^{-1}$) & $1.7 $  \\
SFE                 &Star Formation Efficiency ($L_{\sun}\ M_{\sun}^{-1}$) & $190 $    \\ %
$\tau_{\mathrm{SF}}$&Gas Depletion Lifetime (Myr) & 30 \\
$M_{\mathrm{V}}$             &Optical Absolute Magnitude &  -24.2 \\
$M_{\mathrm{BH}}$   &Black Hole Mass ($10^{9}\ M_{\sun}$)& $1.7$\\
\hline                                   

\end{tabular}
\end{table}

\subsection{J100038+020822: A Starburst-QSO Composite}
The general picture for QSO and stellar spheroid formation is based on the merger of two gas-rich disk galaxies. The large amounts of gas and dust involved in these mergers provide the fuel for infrared luminous starbursts to occur, and the gas inflow into the inner regions of the galaxy likely feeds the central massive black hole \citep{Mihos1994,Barnes1996}. 
Observations of local ULIRGs indicate that the most luminous phase occurs close to the final stage of the merger, when both galaxy disks overlap \citep{Sanders1988a,Sanders1988b,Veilleux1999}.

When the central massive black hole has reached a sufficient size and luminosity, feedback from an active nucleus phase dissipates the dust \citep{DiMatteo2005} and an optically luminous QSO emerges. As the gas is being expelled, the feeding of the QSO ceases, stopping its activity, and the system relaxes into a spheroidal galaxy hosting a (quiescent) supermassive black hole at its center \citep{Hopkins2006}. 

One important question in galaxy evolution is whether this scenario connecting gas-rich starburst galaxies and optically bright QSOs constitutes a rule for most of these systems or whether it is a phenomenon that applies only to the most luminous and massive objects. This evolutionary connection has been studied with vast supporting evidence in the local Universe \citep[e.g.][]{Sanders1988a}. However, for the crucial epoch of star formation and QSO activity ($1<z<3$) only a few studies have so far been undertaken. The study of submillimetre \citep{Page2004, Stevens2005} and CO line \citep{Coppin2008} emission of absorbed and unabsorbed high-redshift QSOs indicates that the QSO phase could be preceded by a SMG starburst phase and suggests that submillimetre selected QSOs represent ideal objects for studying these transitional cases at $z>1$.

J100038 appears to be in a transition between the starburst and QSO phases, as a comparison of its properties to those of a sample of typical SMGs \citep{Greve2005} and QSOs \citep[][]{Solomon2005} shows. We start by describing its QSO properties followed by the features that classify this source as a starburst galaxy.

J100038 shows an optical spectrum typical of a Type-1 AGN (Figure \ref{fig:spectra}) with a prominent MgII broad emission line. It is relatively luminous at optical wavelengths ($M_{\mathrm{V}}\sim-24$ magnitudes, extinction corrected), shows mild optical extinction with $A_{\mathrm{V}}\sim1$ and its X-ray emission indicates it is a heavily absorbed QSO (log$N(\mathrm{H})=22-23$ cm$^{-2}$). Moreover, its optical to mid-IR SED resembles that of typical QSOs (Figure \ref{fig:sed}), well in agreement with \citet{Elvis1994} templates.
In addition, we derived a diameter of $\sim1.5$ kpc for the CO line emitting region. This size is consistent with values observed in high-redshift QSOs which range between 1 and 3 kpc \citep{Walter2004, Solomon2005, Maiolino2007} and similar to those found in local ULIRGs  \citep[$\sim 0.5$ kpc;][]{Downes1998, Soifer2000}, but smaller than the diameters for the CO emitting region in SMGs \citep[$\lesssim4$ kpc;][]{Tacconi2006}.

On the other hand, J100038 shows distinctive features of on-going starburst activity. The optical morphology of J100038 seen in the HST imaging is suggestive of a recent merger event (Figure \ref{fig:closeup_i}). While most of the emission is concentrated in the point-like central source ($C$), the faint emission from the eastern source ($E$) may be indicative of a tidal tail, hinting at a past interaction or merger, although it could also imply the influence of gravitational lensing. The low number density of HST $I$-band sources in the surroundings of J100038 , $\rho(I<25.5)=62$ arcmin$^{-2}$, implies that the probability of chance association between $E$ and $C$ is only 4.1\%. If both sources ($E$ and $C$) are physically related, the projected distance between them would be 7.5 kpc. A large fraction of the optical emission from the host galaxy in J100038 could still be absorbed by surrounding dust (Jahnke et al, in prep.), and could be hiding the actual link to the eastern component. We note that the optical emission of the host galaxy is consistent with the obscured SED of the prototypical local starburst galaxy Arp220. Indeed, the optical morphology of this system is strikingly similar to that observed in local ULIRG/PG-QSOs \citep[IRAS 05189-2524;][]{Sanders1988b, Surace1998}. 
Furthermore, the far-IR to radio SED of J100038 agrees very well with that of Arp220 (Figure \ref{fig:sed}), and does not follow the typical behavior of radio-loud or radio-quiet QSOs at these wavelengths. In fact, its radio to far-IR spectral index, when used as a redshift indicator results in $z=1.9$ \citep{Bertoldi2007}, a strong indication that the far-IR and radio emission are both produced by star formation. As mentioned before, the analysis of the molecular gas physical conditions suggest that the AGN plays a moderate role in the gas heating and hints that the heating may actually be dominated by starforming regions.

All the evidence exposed permits to well fit J100038 in the Sanders et al. scenario. Studies of these transitional cases in the local Universe have found that about $30-50\%$ of the most luminous ULIRGs ($L_{\mathrm{FIR}}>10^{12.3}\ L_{\sun}$) show broad emission lines \citep{Veilleux1999} and the presence of compact nuclei in 40\% of all ULIRGs \citep{Scoville2000,Soifer2000} as well as large molecular gas reservoirs \citep{Evans2002, Evans2005}, even in the late stages of this evolutionary sequence. All these properties apply to J100038. Furthermore, the presence of both starburst and QSO attributes imprinted in the galaxy SEDs have largely been observed in nearby objects, constituting the basis of this evolutionary scenario \citep{Sanders1988a, Sanders1988b, Sanders1989}. These templates compare favorably with the SED observed for J100038, strongly suggesting that this object is evolving from a starburst to a QSO. In fact, it is possible to classify it in a stage between the ``warm ULIRG'' and the ``infrared excess'' QSO phases following the mentioned scenario \citep{Sanders2004}.

\section{Summary}

We detected CO emission from the millimeter selected AGN J100038+020822 at $z=1.8275$ in the COSMOS field. 

J100038 is a millimeter-bright, blank-field selected galaxy classified via its optical spectra as a Type-1 AGN. 
The CO line intensities peak at the $J=5-4$ transition and an LVG analysis of the CO SED finds that the molecular gas in this source is less excited than in typical high-redshift QSOs (e.g. J1148+5251, BR1202-0725 and APM08279+5255), more similar to what is found for SMGs ($T_{\mathrm{kin}}\sim 95$ K, $n(\mathrm{H}_{2})=10^{3.6}$ cm$^{-3}$) such as J16359+6612. 
The CO emission seems to be concentrated in the central kpc, as the comparison between the modeled brightness and observed main beam temperatures shows. The implied molecular gas mass of $(3.6 - 5.4) \times 10^{10} M_{\sun}$ could account for a substancial fraction of the implied dynamical mass within this radius ($<0.75$ kpc). 

Assuming a gray-body far-IR spectrum we derive a dust mass of $1.2 \cdot 10^{9} M_{\sun}$ and a dominant cold dust component with $T_{C}=42\pm4$ K. The broad MgII line allowed us to estimate the central black hole mass to $1.7 \times 10^{9}\ M_{\sun}$.

Although the molecular gas and dust properties are similar to those of typical SMGs and some ULIRGs, the SED from the X-rays to the mid-IR is typical of a QSO. The optical morphology of J100038+020822 is complex, showing evidence of a possible interaction or merger. Its shared properties of starburst and AGN suggest that this composite galaxy fits the evolutionary scenario of starburst to QSO proposed by \citet{Sanders1988a}. 


\begin{acknowledgements}
Manuel Aravena was supported for this research through a stipend from the International Max Planck Research School (IMPRS) for Radio and Infrared Astronomy at the Universities of Bonn and Cologne. 

This work is based on observations with 30m telescope of the \textit{Institute for Radioastronomy at Millimeter Wavelengths} (IRAM), which is funded by the German Max Planck Society, the French CNRS and the Spanish National Geographical Institute. Also based on observations with the Very Large Array of the National Radio Astronomy Observatory, which is a facility of the National Science Foundation, operated under cooperative agreement by Associated Univ. Inc. Based on observations with the NASA/ESA Hubble Space Telescope, obtained at the Space Telescope Science Institute, which is operated by AURA Inc, under NASA contract NAS 5-26555; also based on data collected at: the Subaru Telescope, which is operated by the National Astronomical Observatory of Japan; the XMM-Newton, an ESA science mission with instruments and contributions directly funded by ESA Member States and NASA; the European Southern Observatory, Chile; Kitt Peak National Observatory, Cerro Tololo Inter-American Observatory, and the National Optical Astronomy Observatory, which are operated by the Association of Universities for Research in Astronomy Inc. (AURA) under cooperative agreement with the National Science Foundation; and the Canada-France-Hawaii Telescope operated by the National Research Council of Canada, the Centre National de la Recherche Scientifique de France and the University of Hawaii.

\end{acknowledgements}
\bibliographystyle{aa} 
\bibliography{Aravena_J100038_FB3_editor} 
\clearpage
\end{document}